\documentclass[%
 amsmath,amssymb,
 aps,
]{revtex4-2}
\usepackage{graphicx}
\usepackage{dcolumn}
\usepackage{bm}
\usepackage{textcomp}
\usepackage{booktabs}
\usepackage{float}
\usepackage{makecell}
\usepackage{xcolor}
\newcommand{\be}{\begin{equation}}
\newcommand{\ee}{\end{equation}}
\newcommand{\bea}{\begin{eqnarray}}
\newcommand{\eea}{\end{eqnarray}}

\begin{document}

\preprint{APS/123-QED}

\title{Quantum Correlations in the Decay of $B^0$ meson and Entanglement Entropy}

\author{Divya Sharma}%
 \email{2404110006@mail.jiit.ac.in}
\affiliation{%
Department of Physics and Materials Science and Engineering, Jaypee Institute of Information Technology, Noida, India, 201309 
}%
\author{Vaibhav Rawoot}%
 \email{vaibhav.rawoot@mail.jiit.ac.in}
\affiliation{%
 Department of Physics and Materials Science and Engineering, Jaypee Institute of Information Technology, Noida, India, 201309 
}%
\author{Sudip Kumar Haldar}%
 \email{sudip.haldar@mail.jiit.ac.in}
\affiliation{%
 Department of Physics and Materials Science and Engineering, Jaypee Institute of Information Technology, Noida, India, 201309 
 }%

\date{\today}

\begin{abstract}
We present a phenomenological study of quantum correlations in the decay of $B^0$ mesons into a system of two vector mesons. The decay of the $B^0$ meson into two vector mesons constitutes a bipartite system of two qutrits. The entanglement entropy is used as a measure of quantum correlations in the system of decaying particles. We study the variation of the R\'enyi entropy with R\'enyi order ($\alpha$) for the decay channels $B_s^0 \rightarrow \phi\, \phi$, $B_d^0 \rightarrow J/\psi\, K^{*}(892)^0$, $B_d^0 \rightarrow \phi\, K^{*}(892)^0$ and $B_s^0 \rightarrow J/\psi\, \phi$ and discuss the significance of entanglement entropy at different R\'enyi order regimes. The LHCb, ATLAS and Belle collaborations experimental measurements of complex polarization amplitudes and relative phases are used as input for our analysis. A comparison of entanglement entropy for all the $B^0$ meson decay processes, with both vanishing and non-vanishing phases, reveals a strong phase dependence of the entropy. We further present the results of Hartley entropy (Max-Entropy), von Neumann entropy, collision entropy, and min-entropy, each corresponding to different values and limits of the R\'enyi order. The comparison between the branching fractions of the decay processes and the von Neumann entropy shows a  connection between entanglement and decay dynamics, indicating the role of weak and strong interaction in generating quantum entanglement. In addition, we evaluate several other entanglement measures, including linear entropy, I-concurrence, tangle, negativity, logarithmic negativity, Schmidt coefficients, and Schmidt rank for different $B^0$ meson decay processes. Our study demonstrates that entanglement measures provide useful insights into the underlying decay dynamics and may serve as important tools for understanding quantum correlations in high-energy particle physics processes. 
\end{abstract}
\maketitle
\section{\label{sec:level1}Introduction}
The study of quantum correlations in high-energy particle systems are based on a conceptual framework built from several important contributions to quantum theory. Quantum correlations in physical systems have their conceptual origin in the 1935 thought experiment of Einstein, Podolsky, and Rosen, which questioned the completeness of quantum mechanics through nonlocal correlations in composite systems~\cite{einstein1935can}. In the same year, Schrodinger introduced the term entanglement to describe the phenomenon by which two interacting systems can no longer be described independently~\cite{Schrodinger:2008opj}. The Bell’s theorem, through Bell's inequality, rigorously showed that no local hidden-variable theory can reproduce quantum predictions for entangled particles~\cite{bell1964einstein}. A practically testable version of these inequalities was formulated by Clauser, Horne, Shimony, and Holt, and subsequently confirmed in experiment~\cite{clauser1969proposed,aspect1982experimental}. The Entanglement then became central to quantum information, enabling quantum key distribution and teleportation~\cite{PhysRevLett.70.1895,PhysRevA.53.2046,Vogel01112011}. These foundational concepts underpin quantum information science, where entanglement acts as a critical resource~\cite{RevModPhys.81.865}. Entanglement entropy, typically quantified via the von Neumann entropy of a subsystem's reduced density matrix, provides a fundamental measure of quantum correlations in many-body systems and quantum field theory~\cite{PhysRevA.53.2046,Calabrese:2004eu}.

Quantum entanglement has long been considered mainly a property of low-energy quantum systems such as photons and atoms. However, in recent years, an important question emerged in high-energy physics: can a delicate quantum feature like entanglement survive in extremely energetic particle collisions? To answer this, quantum correlations in various systems in particle physics have been extensively studied in the recent past~\cite{Barr:2024djo,afik2025quantum}. 
  
An important experimental observation in this direction came from the study of top-antitop quark systems at the Large Hadron Collider. The ATLAS collaboration observed quantum entanglement in top quark pair production using proton-proton collision data at $\sqrt{s}=13~\mathrm{TeV}$~\cite{ATLAS:2023fsd}. The top quark is the only quark that does not form a bound state because it decays before the hadronization process can take place. This fast decay causes the spin information of the top quark to be transferred to the final-state decay products. By studying the angular distributions of the final-state leptons, the collaboration observed strong quantum correlations in the $(t\bar t)$ system. Following this result, the CMS Collaboration also investigated quantum entanglement in top-quark pair production \cite{CMS:2024pts}. This collaboration analysis showed that the observed spin correlations cannot be understood only through classical physics. This result further strengthened the idea that heavy particles produced in high-energy collisions can form entangled quantum states.

At the same time, several theoretical studies built a strong connection between collider physics and quantum information theory \cite{Fabbrichesi:2021npl}. The Bell inequalities and entanglement measures were introduced as useful tools for studying quantum correlations in high-energy particle systems. A prominent studies in this direction include the study of quantum entanglement in $\Lambda$-hyperon systems produced in collider experiments \cite{Gong:2021bcp}, charmonium decay system~\cite{Fabbrichesi:2024rec}, $e^+e^- \rightarrow \tau^+\tau^-$ produced in the Belle II experiment \cite{Ehataht:2023zzt} and Higgs boson decays \cite{Barr:2021zcp,Aguilar-Saavedra:2022wam,higgstoleptons}. Together, these developments established collider experiments as an important laboratory for studying quantum entanglement in high-energy particle systems.

The interest in quantum correlations was later extended to heavy meson systems. In particular, $B^0 \rightarrow J/\psi K^{*}(892)^0$ decays were studied from the perspective of polarization entanglement and violation of Bell's inequality~\cite{LHCb:2013vga}. Using experimentally measured helicity amplitudes from the LHCb Collaboration, the spin correlations between the final-state vector mesons were investigated in detail for the decay process $B_s^0 \rightarrow \phi\phi$. Experimentally measured polarization amplitudes and strong phases were used to reconstruct the helicity density matrix of the two vector meson system to study the quantum entanglement and violation of the Bell's inequality. The study revealed a large degree of entanglement in the final state together with the Bell's inequality violation above the $5\sigma$ level. 
The Belle Collaboration investigated the decay in the system of $B_d^0 \rightarrow \phi\, K^{*}(892)^0$\cite{Belle:2005lvd}. The longitudinal and transverse polarization amplitudes were measured, and the analysis showed that transverse polarization contributions are also important in this decay channel.

The ATLAS collaboration studied decay $B_s^0 \rightarrow J/\psi \phi$ using proton-proton collision data collected at the Large Hadron Collider (LHC) \cite{ATLAS:2020lbz}. A detailed angular analysis was performed to measure the polarization amplitudes, the strong phases, and the $CP$-violating phase by the ATLAS collaboration~\cite{fabbrichesi2025measuring}. The measured results were found to be consistent with the predictions of the standard model and later became important inputs for studies of helicity correlations and quantum entanglement in $B^0$-meson decay. 

The measurements of complex polarization amplitudes and phases in various collider experiments enabled us to study quantum correlations in  $B^0$ meson decaying into two systems of vector mesons, as these parameters specify the helicity state of the system. The system of two massive vector mesons is important for studying entanglement in the qutrit system. The bound entanglement in the two qutrit state and the separability of the joint $N$ qutrit are relevant for quantitative studies related to bipartite entanglement in higher-dimensional systems~\cite{CAVES2000439, PhysRevA.94.020302}. 

In this paper, we have shown the variation of R\'enyi entropy for both vanishing and non-vanishing phases of states of the system. The measurement of complex polarization amplitudes and phases from collider experiments has been used for calculating the R\'enyi entropy. We also estimate errors in the  R\'enyi entropy resulting from errors in measurements of amplitude and phases. We also show the relation between branching fractions for decay processes and the corresponding von Neumann entropy. Our study is relevant in understanding the nature of entanglement entropy for different values and the limit of the R\'enyi order parameter $\alpha$. The variation of R\'enyi entropy with R\'enyi order is studied for the first time for the system of elementary particles. Also we calculate different types of entanglement measures, such as linear entropy $(S_L)$, I-concurrence $(C_I)$, tangle $(\tau)$, negativity $(\mathcal{N})$, logarithmic negativity $(E_{\mathcal{N}})$, Schmidt coefficients and Schmidt rank for different $B^0$ meson decay processes~\cite{joshi2026entanglement}.

The article is organized as follows. In Section II, we give the formalism of the density matrix, the helicity states of the system, and the relation between helicity amplitudes and polarization amplitudes for the decay of \(B^0\) mesons into two vector mesons. Using the reduced density matrix formalism, we study quantum correlations through the R\'enyi entanglement entropy for different values and limits of the R\'enyi order \(\alpha\). We also discuss the formalism of the von Neumann entropy, Hartley entropy, collision entropy, min-entropy, linear entropy, I-concurrence, tangle, negativity, logarithmic negativity, Schmidt coefficients, and Schmidt rank for the two vector meson system. In Section III, we present the numerical results and plots for different \(B^0\) meson decay processes using polarization amplitudes and relative phases from the LHCb, Belle, and ATLAS collaborations. We study the variation of R\'enyi entropy with the R\'enyi order \(\alpha\) and also compare the results for vanishing and non-vanishing phases. We further calculate the entanglement entropy and compare the  branching fractions and the von Neumann entropy for different decay processes. We also show the results of several other entanglement measures, including linear entropy, I-concurrence, tangle, negativity, logarithmic negativity, Schmidt coefficients and Schmidt rank for different \(B^0\) meson decay processes to understand the quantum correlation in the decay of $B^0$ meson. We summarize our results and draw conclusions in Section $\text{IV}$.

\section{Formalism}
In this section, we will describe the basic formalism required to study quantum correlations in the system of particles. The system of our interest is the two massive vector bosons decaying from spin-0 particle. The massive vector boson has three polarization states. The polarization states of the particle can also be expressed in terms of helicity states~\cite{Fabbrichesi:2023idl}. The two-particle quantum state can be described by the two-party density matrix, which can be used to define different measures of quantum correlation.

\subsection{Density Matrix}
The density matrix is an important formalism for studying quantum correlations in the system of particles~\cite{blum2012density}. A quantum system can be in the pure state or in the mixed state. The density matrix for the pure state is given by
\begin{equation}
    \rho=|\psi\rangle\langle\psi|
    \label{eq:1}
\end{equation}
Here, $|\psi\rangle$ represents the state of the quantum system.
An ensemble $\{(p_i,|\psi_i\rangle)\}$ of pure states $|\psi_i\rangle$ describes the mixed system. The Density matrix for the mixed state is given by
\begin{equation}
    \rho=\sum_i p_i|\psi_i\rangle\langle\psi_i|, \,\,\,\,\,\,\,\text{with}\,\,\,\, p_i\geq0 \,\,\,\,\,\,\, \sum_i p_i =1
    \label{eq:2}
\end{equation}
The following are the properties of the density operator (density matrix):
\begin{enumerate}
    \renewcommand{\labelenumi}{(\roman{enumi})}
    \item Trace: $ \mathrm{Tr}(\rho)=1 $
    \item Hermiticity: $ \rho=\rho^\dagger $
    \item Positivity: $ \rho \geq 0 $
\end{enumerate}

The pure and mixed states can be identified using the density operator. For pure states, $\rho^2 = \rho$, which implies that the 
$Tr(\rho^2) = 1$, whereas for the mixed state, $Tr(\rho^2) \leq 1$.

\subsection{Helicity State of the System}
To describe the decay of $B^0$ meson into the system of two massive vector mesons, we introduce the helicity amplitudes. The helicity amplitudes are expressed in terms of the interaction Hamiltonian $\mathcal{H}_I$, which controls the decay of the $B^0$ meson. This Hamiltonian includes the interactions that cause the decay that modify the final state. The transition amplitude for the decay of the $B^0$ meson into two vector mesons $V_1$ and $V_2$ is written as

\begin{equation}
w_{\lambda \lambda'} \propto \langle V_1(\lambda)\, V_2(-\lambda') \,|\,\mathcal{H}_I \,|\, B \rangle
\label{eq:3}
\end{equation}
with, $\lambda = \lambda' = +, 0$ and -, represent the three helicity states of massive vector mesons. The interaction hamiltonian $\mathcal{H}_I$ carries the dynamics of the transition amplitude~\cite{fabbrichesi2023bell}. Hence, the $B^0$ meson decays into two vector mesons with definite spin directions. The $B^0$ meson is a spin-0 particle, and both vector mesons in the final state have spin-1. Therefore, the total angular momentum must remain conserved during the decay process. The conservation laws allow only three configurations of two particle states. Now we introduce three amplitudes, $w_{++}, w_{00}$ and $ w_{--}$, which describe the final state in the decay. Its magnitude caries information about probability of each configuration. The final state is a combination of all three allowed states. So, the quantum state is written as
\begin{equation}
|\psi\rangle =
\frac{1}{\sqrt{|\mathcal{M}|^2}}
\left(
w_{++}|++\rangle
+
w_{00}|00\rangle
+
w_{--}|--\rangle
\right),
\label{eq:4}
\end{equation}
where
\begin{equation}
|\mathcal{M}|^2 =
|w_{++}|^2 + |w_{00}|^2 + |w_{--}|^2.
\label{eq:5}
\end{equation}
This normalization ensures that the total probability is equal to 1.
The two-massive-vector-meson system presents a interesting scenario in which all three spin configurations contribute simultaneously. The relative magnitudes of the amplitudes determine the dominant configuration, while their phases govern the interference pattern. Consequently, the two final-state particles cannot be treated as independent, their spin degrees of freedom are quantum-mechanically correlated prior to the decay.
To study the system more carefully, we express it in a complete basis. Each massive vector meson has a three helicity states $|+\rangle$, $|0\rangle$, and $|-\rangle$. So, the combined system has nine basis states, 
$|+-\rangle,\ |+0\rangle,\ |++\rangle,\ 
|0+\rangle,\ |00\rangle,\ |0-\rangle,\ 
|-+\rangle,\ |-0\rangle$ and $\ |--\rangle.$
This gives a 9-dimensional Hilbert space. But our physical state does not occupy all 9-dimensional Hilbert space. Because the initial $B^0$ meson is spin-0 particle, angular momentum conservation allows only three components to be non-zero. The possible representation of this basis in vector form is
\begin{equation}
\begin{aligned}
|+-\rangle &= \begin{pmatrix} 1 \\ 0 \\ 0 \\ \vdots \\ 0 \end{pmatrix}, \quad 
|+0\rangle = \begin{pmatrix} 0 \\ 1 \\ 0 \\ \vdots \\ 0 \end{pmatrix}, \\
|++\rangle &= \begin{pmatrix} 0 \\ 0 \\ 1 \\ \vdots \\ 0 \end{pmatrix}, \quad 
\dots \quad 
|-+\rangle = \begin{pmatrix} 0 \\ 0 \\ 0 \\ \vdots \\ 1 \end{pmatrix}
\end{aligned}
\label{eq:6}
\end{equation}

In matrix form, each basis state is a column vector with only one non-zero entry. When we write the full state $|\psi\rangle$, it becomes a three dimensional vector. The only allowed basis are $|++\rangle$, $|00\rangle$, and $|--\rangle$~\cite{Gabrielli:2024kbz}.

The two massive vector meson system, though relatively simple, exhibits strong quantum correlations. The spins of the two particles are correlated through the spin and angular momentum conservation laws governing the decay process itself. This makes the decay of the $B^0$ meson into two massive vector mesons a particularly clean and suitable process for studying quantum correlations at high energy collider experiments, where spin and angular momentum conservation directly give rise to entanglement. The density matrix of the system is constructed in the helicity basis, from which the various measures of quantum entanglement are calculated.

\subsection{Density Matrix for the System}
Now we consider the density matrix formalism to study quantum correlations in the system of massive vector mesons~\cite{Gabrielli:2024kbz}. The outer product of the quantum state given in equation~\eqref{eq:1} gives the $9 \times 9$ density matrix. We obtain the density matrix using the basis vectors given in equation~\eqref{eq:6} and the quantum state given in equation~\eqref{eq:4} as
\begin{equation}
\
\rho =
\frac{1}{|M|^2}
\begin{pmatrix}
0 & 0 & 0 & 0 & 0 & 0 & 0 & 0 & 0\\
0 & 0 & 0 & 0 & 0 & 0 & 0 & 0 & 0\\
0 & 0 & w_{++}w_{++}^{*} & 0 & w_{++}w_{00}^{*} & 0 & w_{++}w_{--}^{*} & 0 & 0\\
0 & 0 & 0 & 0 & 0 & 0 & 0 & 0 & 0\\
0 & 0 & w_{00}w_{++}^{*} & 0 & w_{00}w_{00}^{*} & 0 & w_{00}w_{--}^{*} & 0 & 0\\
0 & 0 & 0 & 0 & 0 & 0 & 0 & 0 & 0\\
0 & 0 & w_{--}w_{++}^{*} & 0 & w_{--}w_{00}^{*} & 0 & w_{--}w_{--}^{*} & 0 & 0\\
0 & 0 & 0 & 0 & 0 & 0 & 0 & 0 & 0\\
0 & 0 & 0 & 0 & 0 & 0 & 0 & 0 & 0
\end{pmatrix}.
\label{eq:7}
\end{equation}

The diagonal elements of the density matrix are given by $|w_{++}|^2$, $|w_{00}|^2$, and $|w_{--}|^2$. These represent the probabilities of finding the system in each helicity configuration.

We begin by identifying the system $B^0 \to V_1V_2$ as a bipartite quantum system that lives in a composite Hilbert space. One vector meson is labeled particle $A$ and lives in Hilbert space $\mathcal{H}_A$, while the other vector meson is labeled particle $B$ and lives in Hilbert space $\mathcal{H}_B$. Therefore, the total Hilbert space of the composite system can be written as
\begin{equation}
\mathcal{H}_{AB} = \mathcal{H}_A \otimes \mathcal{H}_B.
\label{eq:8}
\end{equation}

This naturally leads to the concept of partial trace. Instead of describing the full system, one traces over the unobserved degrees of freedom to obtain a reduced density matrix that encodes only the accessible subsystem. The partial trace acts as a linear operation that maps the density matrix of the composite system $\mathcal{H}_{AB}$ to the reduced density matrix of a subsystem $\mathcal{H}_A$ or $\mathcal{H}_B$.
We construct an effective description that lives entirely in the Hilbert space $\mathcal{H}_A$ or $\mathcal{H}_B$.

So, the reduced density matrix of the particle $A$ can be written as
\begin{equation}
\rho_A = \mathrm{Tr}_B(\rho).
\label{eq:9}
\end{equation}
Now in this bipartite system, only three helicity states are allowed which are $|++\rangle$, $|00\rangle$, and $|--\rangle$. When we perform the trace over particle $B$, all off-diagonal terms vanish while only the diagonal terms remain. Taking a partial trace over system $B$, we obtain the reduced density matrix as 
\begin{equation}
    \rho_A =
\frac{1}{|M|^2}
\begin{pmatrix}
|w_{++}|^2 & 0 & 0 \\
0 & |w_{00}|^2 & 0 \\
0 & 0 & |w_{--}|^2
\end{pmatrix}.
\label{eq:10}
\end{equation}

The diagonal elements give the probabilities of finding the particle $A$ in different helicity states. The reduced density matrix shows that particle $A$ cannot be described independently. Its properties depend on particle $B$, even if particle $B$ is not observed.

So, the reduced density matrix describe how a full quantum system reduces to measurable quantities, while still keeping the signature of quantum correlations between the two particles. The diagonal elements of the reduced density matrix are its eigenvalues. From equation~\eqref{eq:10}, these are given by
\begin{equation}
\lambda_1 = \frac{|w_{++}|^2}{|M|^2}, \quad
\lambda_2 = \frac{|w_{00}|^2}{|M|^2}, \quad
\lambda_3 = \frac{|w_{--}|^2}{|M|^2}.
\label{eq:11}
\end{equation}

These eigenvalues contain complete information about the system at the level of one particle. The probability is distribution among the different helicity states is described by eigenvalues. If one eigenvalue is very large compared to the others, it means that one helicity configuration dominates the decay. On the other hand, if all three eigenvalues are similar, then all configurations contribute almost equally. The distribution of these eigenvalues directly reflects the strength of the quantum correlations between the two particles. 

\subsection{Polarization Amplitude and Helicity Amplitude}
 We consider helicity amplitudes, $w_{++}, w_{00}, $ and $ w_{--}$, which survives naturally from angular momentum conservation. The formalism of the helicity amplitude is correct and complete, but it is not always the easiest to use when we want to connect with experiments.
 
Here we introduce the polarization amplitudes $A_0, A_{\parallel}$ and $ A_{\perp}$ of the decaying vector meson. The amplitude $A_0$ describes the longitudinal polarization~\cite{Gabrielli:2024kbz}. The amplitudes $A_{\parallel}$ and $A_{\perp}$ describe transverse polarizations.
The relation with helicity amplitudes is given by 
\begin{equation}
w_{++} = \frac{A_{\parallel} + A_{\perp}}{\sqrt{2}}, \quad
w_{--} = \frac{A_{\parallel} - A_{\perp}}{\sqrt{2}}, \quad
w_{00} = A_0.
\label{eq:12}
\end{equation}

Each polarization amplitude is a complex number and can be written as
\begin{equation}
A_i = |A_i| e^{i\delta_i}.
\label{eq:13}
\end{equation}
The magnitude $|A_i|$ quantifies the relative contribution of each polarization amplitude, while the phase $\delta_i$ carries information about the dynamics of the decay. The reference phase is usually chosen as $\delta_0 = 0$. This leaves two independent phases, $\delta_{\parallel}$ and $\delta_{\perp}$. 
The phases of the helicity amplitudes are of particular importance, as they dictate how the different polarization components combine coherently in the full quantum state, directly influencing the interference structure and the resulting entanglement between the two decay products.

The squared magnitudes $|A_0|^2$, $|A_{\parallel}|^2$, and $|A_{\perp}|^2$ represent the polarization fractions. The condition,
\begin{equation}
|A_0|^2 + |A_{\parallel}|^2 + |A_{\perp}|^2 = 1, 
\label{eq:14}
\end{equation}
ensures the normalization.

\subsection{Entanglement Entropy}
Quantifying entanglement in a quantum system is a nontrivial task, and the complexity of the problem grows significantly with the dimensionality of the system. For two-level systems, such as qubits, many standard tools are available. However, when we move to higher-dimensional systems, like massive spin-1 particles in our case, the problem becomes more challenging. In such situations, a single measure is often not sufficient, and different quantities such as R\'enyi entropy, max-entropy, von Neumman entropy, collision entropy and min-entropy are used to capture different aspects of entanglement. The entropy of entanglement is one of the important measure of quantum correlations in the higher dimensional system.

The eigenvalues $\lambda_1, \lambda_2$ and $\lambda_3$ of the reduced density matrix contains information about the system. Thus, we can study entanglement directly using these eigenvalues instead of working with the full quantum state. 
To quantify the entanglement in quntum system, we employ the R\'enyi entropy~\cite{ozawa2024perspective}, a generalized from the von Numann entropy. It is defined as
\begin{equation}
S_\alpha = \frac{1}{1 - \alpha} \log \left( \sum_i \lambda_i^\alpha \right).
\label{eq:15}
\end{equation}
Here, $\lambda_i$ are the eigenvalues and $\alpha$ is the R\'enyi order. The eigenvalues $\lambda_i$ describe how the probability is distributed among different helicity states. We use natural logarithm in our analysis, denoted by $\log(x)$.

\begin{figure}[H]
\centering
\includegraphics[width=0.5\textwidth]{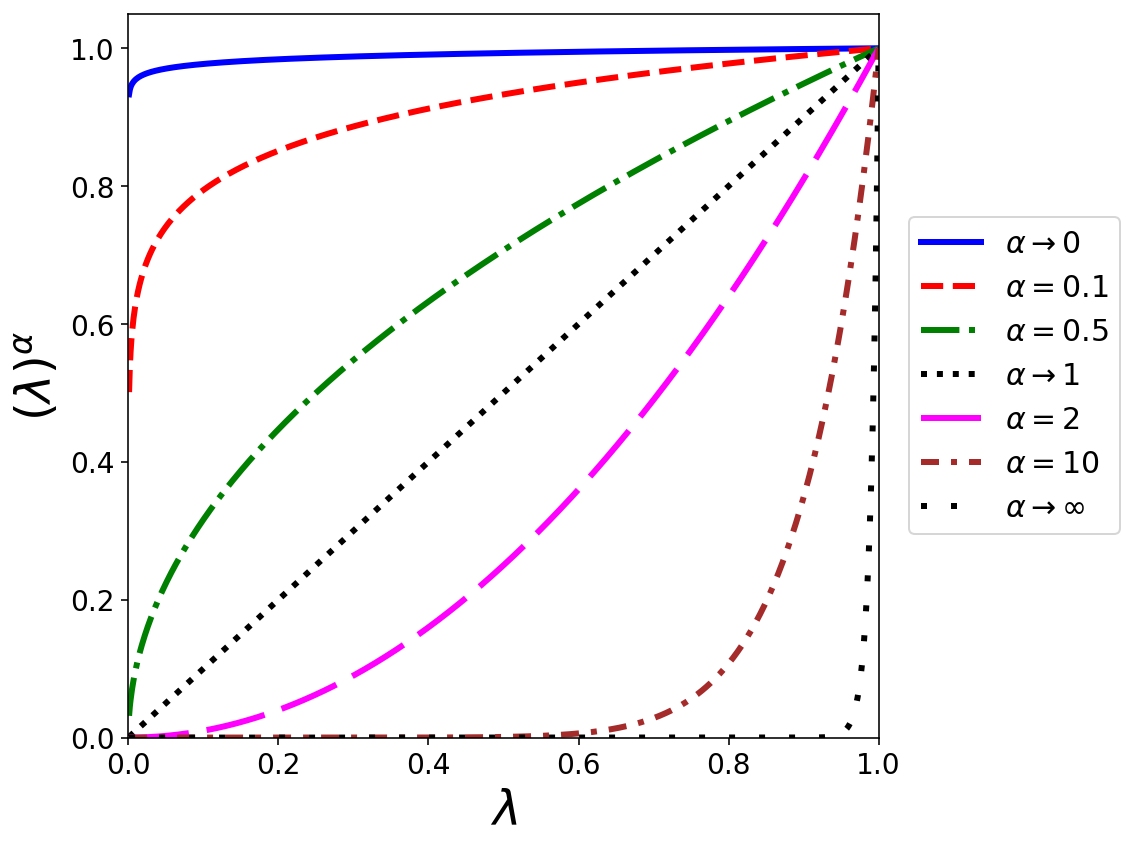}
\caption{(Color online) Behavior of the function $\lambda^\alpha$ for different values of the Rényi order $\alpha$. The solid blue curve shows $\alpha \to 0$. The red dashed curve shows $\alpha = 0.1$. The green dash-dotted curve shows $\alpha = 0.5$. The black dotted straight line shows $\alpha \to 1$. The magenta long-dashed curve shows $\alpha = 2$. The brown dash-dotted curve shows $\alpha = 10$. The black sparsely dotted curve shows $\alpha \to \infty$.}
\label{fig:1}
\end{figure}

The parameter $\alpha$ plays a central role in shaping the sensitivity of the entropy measure. For small $\alpha$, all eigenvalues contribute almost equally. In this region, the entropy tells us how many helicity states are active in the decay. As $\alpha$ increases, larger eigenvalues become more important. This makes the entropy sensitive to the dominant helicity configuration. 
The eigenvalues of the reduced density matrix are determined by the helicity amplitudes, which are fixed by the underlying decay dynamics. Entanglement is therefore encoded in the structure of the decay itself, rather than being an independent feature.
It shows that the entanglement is already present in the structure of the decay itself. The R\'enyi order and its weighing of eigen values enable us to introduce the different regions of $\alpha$.
The R\'enyi order $\alpha$ and its weighting of the eigenvalue spectrum provide a natural way to define distinct regions of $\alpha$, each probing a different aspect of the helicity structure of the bipartite system.

Figure~\ref{fig:1} shows the behavior of the function $\lambda^{\alpha}$ for different values and limits of the R\'enyi order $\alpha$. The shape of the curves shows different behavior because different values of $\alpha$ treat probabilities in different ways. For $\alpha < 1$, the upper curves bend upward. A very small value of $\lambda$ does not rapidly approach zero after the transformation. Therefore, weak quantum states still contribute to the entropy. For $\alpha \to 1$, the curve becomes a straight line. In this case, the probabilities remain unchanged, which corresponds to the von Neumann entropy limit.
For $\alpha > 1$, the lower curves bend downward. Small probabilities now become much smaller and the largest eigenvalues contribute. In the following subsection, we will describe the entropy associated with the different regions of R\'enyi order $\alpha$~\cite{joshi2026entanglement}.

\subsubsection{Von Neumann Entropy ($\alpha \to 1$)}

The expression for R\'enyi entropy  is undefined at $\alpha = 1$ and must be evaluated using the limit $\alpha \to 1$. In this limit, it reduces to the von Neumann entropy~\cite{ozawa2024perspective}, given by
\begin{equation}
S_1 = - \sum_i \lambda_i \log \lambda_i.
\label{eq:16}
\end{equation}

The von Neumann entropy serves as a standard measure of entanglement in quantum mechanics. It describes how evenly the decay is shared among different helicity configurations. In this way, a simple quantity like entropy connects directly to the physical behavior of the decay process and the structure of the entanglement in the system.

\subsubsection{Max-entropy, Hartley Entropy ($\alpha \to 0$)}
Now we consider the limit $\alpha \to 0$ in the R\'enyi entropy. In this limit, the R\'enyi entropy reduces to the Hartley entropy~\cite{ozawa2024perspective}, also called the max-entropy. In the limit $\alpha \to 0$, the expression becomes
\begin{equation}
S_0 = \log(\mathrm{rank}(\rho_A)) = \log \Omega,
\label{eq:17}
\end{equation}
where $\Omega$ is the number of non-zero eigenvalues.
The Hartley entropy does not depend on the actual values of the eigenvalues. It depends only on the number of non-zero eigenvalue. It therefore counts the number of independent states effectively contributing to the system. Even if all three eigenvalues are non-zero, the entanglement can still be weak if one eigenvalue dominates over the others. In our decay process, the Hartley entropy reflects how many helicity configurations are allowed by the dynamics. If more helicity states are active, the system has more freedom to develop quantum correlations.

\subsubsection{Collision Entropy ($\alpha = 2$)}
Now we consider the case $\alpha = 2$ in the Rényi entropy. In this case, the expression for entropy is given by
\begin{equation}
S_2 = - \log \left( \sum_i \lambda_i^2 \right).
\label{eq:18}
\end{equation}

The quantity $\sum_i \lambda_i^2$ is known as the purity of the reduced density matrix~\cite{Bosyk:2011bay}. It measures the degree to which the subsystem approximates a pure state. Because the eigenvalues appear squared, larger eigenvalues are emphasized more strongly. If one helicity amplitude dominates, the purity is high and the entropy is low. If several amplitudes contribute with similar strength, the system becomes more mixed, the purity decreases, and the entropy increases. The collision entropy is connected to the decay dynamics.

\subsubsection{Min-Entropy ($\alpha \to \infty$)}
We now consider the limit of the R\'enyi entropy when $\alpha \to \infty$, which defines the min-entropy~\cite{ozawa2024perspective}. In this limit, the expression for entropy is given by
\begin{equation}
S_{\infty} = -\log(\lambda_{\text{max}}),
\label{eq:19}
\end{equation}
where $\lambda_{\text{max}}$ is the largest eigenvalue.
The min-entropy depends only on a single quantity, which is the largest eigenvalue. In this limit, only the most probable helicity configuration contributes, while all others are suppressed. 
The min-entropy quantifies the degree to which a single helicity configuration dominates the decay. It is determined solely by the largest eigenvalue of the reduced density matrix and therefore reflects how strongly one helicity channel controls the final state. The min-entropy suppresses smaller contributions and highlights only the dominant one. 

In the next section, we describe other relevant entanglement measures for the bipartite system.

\subsection{Linear Entropy}
The linear entropy is one of the most accessible measures of mixedness and entanglement in a quantum system.~\cite{joshi2026entanglement}. It measures how mixed the reduced density matrix becomes after tracing out one part of a bipartite system. The Linear entropy is defined in terms of reduced density matrix $\rho_A$ as
\begin{equation}
S_L = 1-\mathrm{Tr}(\rho_A^2).
\label{eq:20}
\end{equation}

If the reduced density matrix corresponds to a pure state, then $\rho_A^2=\rho_A$, and therefore $S_L=0$.
This means that the subsystem is completely pure and there is no entanglement between the two particles. However, when the subsystem becomes mixed, the value of linear entropy increases. For the $d$-dimensional subsystem, the maximum limit of the linear entropy is $(1-\frac{1}{d})$. Therefore, the maximum value of the linear entropy for a two qutrit system becomes $2/3$. Thus, larger values of linear entropy indicate stronger quantum correlations and stronger entanglement.

The linear entropy can also be written in terms of the eigenvalues $\lambda_i$ of the reduced density matrix as,
\begin{equation}
S_L
=
1-
\sum_i \lambda_i^2.
\label{eq:21}
\end{equation}

\subsection{I-concurrence}
Concurrence is another important quantity that quantify bipartite entanglement~\cite{rungta2003concurrence}. For three level quantum systems, the generalized form known as the I-concurrence, which is written as 
\begin{equation}
C_I=\sqrt{2(1-\mathrm{Tr}(\rho_A^2))}.
\label{eq:22}
\end{equation}

A larger value of I-concurrence corresponds to stronger entanglement between the two subsystems. When the I-concurrence approaches zero, the state becomes separable. When I-concurrence becomes large, the system shows strong quantum correlations.
The maximum I-concurrence for a bipartite system of dimension $d\times d$ is
{\begin{equation}
C_{I{\mathrm{max}}}
=
\sqrt{\frac{2(d-1)}{d}},
\label{eq:23}
\end{equation}}
where $d=3$ for qutrit system.
Therefore, the maximum limit of I-concurrence for qutrit system is  $\frac{2}{\sqrt{3}}$~\cite{rungta2003concurrence}.

\subsection{Tangle}
The tangle is an important quantity that describes the degree of entanglement present in a bipartite quantum system~\cite{rungta2003concurrence}. It is directly related to I-concurrence and provides a more sensitive description of entanglement distribution in composite quantum systems. The tangle is defined as the square of the I-concurrence,

\begin{equation}
\tau = C_I^2,
\label{eq:24}
\end{equation}
where \(C_I\) represents the I-concurrence of the quantum state. The tangle varies from $\tau=0$ for a product state to $2(d-1)/d$ for a maximally entangled state. For a qutrit system, then the upper limit of $\tau$ becomes $4/3$~\cite{rungta2003concurrence}.
Whereas for $\tau=0$, the subsystems in a product state behave independently and no correlation is shared between them. In the maximally entangled state, the quantum correlation between the subsystems becomes strongest.

\subsection{Negativity and Logarithmic Negativity}
Negativity and logarithmic negativity are prominent measures that quantify the quantum entanglement in the system of particle~\cite{ghosh2019entanglement}. The fundamental premise of negativity is rooted in properties of the partially transposed density matrix. The evaluation of the eigenvalues of the partially transposed density matrix determines if the correlations present within the system are purely classical or genuinely quantum mechanical~\cite{joshi2026entanglement}.

For a bipartite density matrix $\rho$, the negativity is defined as
\begin{equation}
\mathcal{N}(\rho)=\sum_{\lambda_i<0} |\lambda_i|,
\label{eq:25}
\end{equation}
where $\lambda_i$ are the negative eigenvalues of the partially transposed density matrix $\rho^{T_B}$. The presence of these negative eigenvalues clearly shows the signature of quantum entanglement within the system.
Whereas negativity value of zero indicates the absence of any negative eigenvalues, and the system behaves like a separable state. In that scenario, the two vector mesons are no longer quantum mechanically entangled. On the other hand, larger values of negativity indicate stronger quantum correlations between the two vector mesons. The bipartite qutrit negativity has also been experimentally studied in spatial-bin photonic qutrit systems~\cite{ghosh2019entanglement}.
For a maximally entangled bipartite state of dimension $d\times d$, the maximum negativity is given by
\begin{equation}
\mathcal{N}_{\mathrm{max}}
=
\frac{d-1}{2}.
\label{eq:26}
\end{equation}

Therefore, for a two-qutrit ($d\times d$) system, with $d=3$ gives $\mathcal{N}_{\mathrm{max}}=1.$
Hence, the negativity for a two-qutrit system satisfies $0 \leq \mathcal{N}\leq 1.$

The logarithmic negativity is constructed from negativity and is defined as
\begin{equation}
E_\mathcal{N}(\rho)=\log\left(1+2N(\rho)\right).
\label{eq:27}
\end{equation}
This quantity gives another way to quantify the entanglement~\cite{joshi2026entanglement}. While negativity tells us whether entanglement exists, logarithmic negativity also provides a convenient measure of how much entanglement is present in the system.
For a maximally entangled state in a $d\times d$ system,
\begin{equation}
E_{\mathcal{N}}^{\mathrm{max}}
=
\log (d).
\label{eq:28}
\end{equation}

Therefore, for a two-qutrit ($3\times3$) system, $E_{\mathcal{N}}^{\mathrm{max}}
=1.098612$ and the logarithmic negativity for a two-qutrit system lies in the range
$0 \leq E_{\mathcal{N}} \leq \log(3)$.

\subsection{Schmidt Coefficients and Schmidt rank}
The Schmidt decomposition provides a simple and powerful way to describe bipartite quantum states~\cite{Vogel01112011}. Any bipartite pure state can be written as
\begin{equation}
|\psi\rangle
=
\sum_i
\sqrt{s_i}
\,|u_i\rangle_A
\otimes
|v_i\rangle_B,
\label{eq:29}
\end{equation}
where $s_i$ are called Schmidt coefficients and satisfy the normalization condition $\sum_i s_i =1$. Here, \( |u_i\rangle_A \) and \( |v_i\rangle_B \) represent orthonormal basis states belonging to subsystems \(A\) and \(B\), respectively. In the $B^0$ decay to two vector mesons system, the eigenvalues of the reduced density matrix are directly related to the Schmidt coefficients through $\lambda_i = s_i^2$. The Schmidt rank is a useful quantity for studying quantum entanglement in a bipartite system. The Schmidt rank is defined as the number of nonzero Schmidt coefficients appearing in the Schmidt decomposition of the state. Then the Schmidt rank is equal to the total number of nonzero coefficients \(s_i\). Therefore, the maximum limit of the Schmidt rank is 3 for a maximally entangled state of a qutrit system.

The density matrix for the two qutrit system gives the theoretical results for quantum correlations in the system of particles. 
We can use this formalism to do the phenomenological study related to quantum entanglement entropy in the $B^0$ decay to two vector mesons system.
The polarization amplitudes and phases are measured by collider experiments. We use these amplitudes and phases to obtain the quantum states of two massive vector mesons. We use the R\'enyi entropy to measure entanglement and plot the graph between different R\'enyi orders and R\'enyi entropy. After that we calculate the error propagation from the experimental data for different processes. We also compare the results and plots of all the processes to know the variation of R\'enyi entropy. In the next section we discuss our results.

\section{Results and Discussion}
In this work, we study the quantum correlations in $B^0$ meson decays into a system of two massive vector mesons. We calculate the R\'enyi entanglement entropy for different values of the R\'enyi order ($\alpha$). Additionally, we have obtained plots illustrating the variation of entanglement entropy with R\'enyi orders. We carried out the analysis for $B_s^0 \rightarrow \phi\, \phi$ ~\cite{Gabrielli:2024kbz}, $B_d^0 \rightarrow J/\psi\, K^{*}(892)^0$ ~\cite{LHCb:2013vga}, $B_d^0 \rightarrow \phi\, K^{*}(892)^0$~\cite{Belle:2005lvd} and $B_s^0 \rightarrow J/\psi\, \phi$ decay processes~\cite{ATLAS:2020lbz}. We also compare the von Neumann entropy with branching fractions of each decay process. 
In addition, we provide the results of other entanglement measures such as, linear entropy, I-concurrence, tangle, negativity, logarithmic negativity and Schmidt coefficients for different $B^0$ meson decay processes and provide the comparative analysis. 

We provide the data of polarization amplitudes and phases measured by various experimental collaborations in the next subsections.

\subsection{ Polarization amplitudes and relative phases data}
We now present the data of polarization states of massive vector mesons required to calculate entanglement entropy. We used polarization amplitude data from LHCb, ATLAS, Belle and CMS experiment. Using this data, we calculate the density matrix for the system of two quitrits. 

\subsubsection{$B_s^0 \rightarrow \phi\, \phi$}
The polarization amplitudes and the relative phases for the process $B_s^0 \rightarrow \phi\, \phi$ are experimentally measured by LHCb experiments~\cite{Gabrielli:2024kbz}. In Table~\ref{tab:table1}, we present the data of polarization amplitudes and relative phases. Data for correlations in polarization amplitude and relative phases are given in Table~\ref{tab:table2}.

\begin{table}[htbp]
\begin{tabular}{cc}
\toprule
Parameter & Values \\
\midrule
$|A_0|^2$          & $0.384 \pm 0.007 \,(\mathrm{stat.})\pm0.003\,(\mathrm{syst.})$ \\
$|A_{\perp}|^2$    & $0.310 \pm 0.006 \,(\mathrm{stat.})\pm 0.003\,(\mathrm{syst.})$ \\
$\delta_{\parallel}\,[\mathrm{rad}]$ & $2.463 \pm 0.029 \,(\mathrm{stat.})\pm 0.009\,(\mathrm{syst.})$ \\
$\delta_{\perp}\,[\mathrm{rad}]$     & $2.769 \pm 0.105 \,(\mathrm{stat.})\pm 0.011\,(\mathrm{syst.})$ \\
\bottomrule
\end{tabular}
\centering
\caption{The values of polarization amplitudes and relative phases for $B_s^0 \rightarrow \phi\, \phi$. Data is taken from the LHCb experiment collaboration~\cite{Gabrielli:2024kbz}.} 
\label{tab:table1}
\end{table}
\begin{table}[htbp]
\centering

\begin{tabular}{ccccc}
\toprule
 & $|A_0|^2$ & $|A_{\perp}|^2$ & $\delta_{\parallel}$ & $\delta_{\perp}$ \\
\midrule
$|A_0|^2$ & 1 & -0.342 & -0.007 & 0.064 \\
$|A_{\perp}|^2$ &  & 1 & 0.140 & 0.088 \\
$\delta_{\parallel}\,[\mathrm{rad}]$ &  &  & 1 & 0.179 \\
$\delta_{\perp}\,[\mathrm{rad}]$ &  &  &  & 1 \\
\bottomrule
\end{tabular}
\caption{Correlation matrix for the polarization amplitudes and phases for $B_s^0 \rightarrow \phi\, \phi$. Data is taken from the LHCb experiment collaboration~\cite{Gabrielli:2024kbz}.}
\label{tab:table2}
\end{table}

\subsubsection{$B_d^0 \rightarrow J/\psi\, K^{*}(892)^0$ }

The polarization amplitudes and the relative phases for the process $B_d^0 \rightarrow J/\psi\, K^{*}(892)^0$  are experimentally measured by LHCb Experiment collaboration~\cite{LHCb:2013vga}. In Table~\ref{tab:table3}, we present the data of polarization amplitudes and relative phases. Data for correlations in polarization amplitude and relative phases are given in Table~\ref{tab:table4}.

\begin{table}[htbp]

\begin{tabular}{cc}
\toprule
Parameter & Values \\
\midrule
$|A_{\parallel}|^2$ & $0.227 \pm 0.004 \,(\mathrm{stat.}) \pm 0.011 \,(\mathrm{syst.})$ \\

$|A_{\perp}|^2$ & $0.201 \pm 0.004 \,(\mathrm{stat.}) \pm 0.008 \,(\mathrm{syst.})$ \\

$\delta_{\parallel}$ [rad] & $-2.94 \pm 0.02 \,(\mathrm{stat.}) \pm 0.03 \,(\mathrm{syst.})$ \\

$\delta_{\perp}$ [rad] & $2.94 \pm 0.02 \,(\mathrm{stat.}) \pm 0.02 \,(\mathrm{syst.})$ \\
\bottomrule
\end{tabular}
\centering
\caption{The values of polarization amplitudes and relative phases for $B_d^0 \rightarrow J/\psi\, K^{*}(892)^0$. Data is taken from the LHCb experiment collaboration~\cite{LHCb:2013vga}.}

\label{tab:table3}
\end{table}

\begin{table}[htbp]
\centering
\begin{tabular}{ccccc}
\hline
 & $|A_{\parallel}|^2$ & $|A_{\perp}|^2$ & $\delta_{\parallel}$ & $\delta_{\perp}$ \\
\hline
$|A_{\parallel}|^2$ & 1.00 & -0.70 & 0.12 & 0.04 \\
$|A_{\perp}|^2$     &      & 1.00  & -0.14 & -0.01 \\
$\delta_{\parallel}\,[\mathrm{rad}]$&      &       & 1.00  & 0.64 \\
$\delta_{\perp}\,[\mathrm{rad}]$    &      &       &       & 1.00 \\
\hline
\end{tabular}
\caption{Correlation matrix for the polarization amplitudes and phases for $B_d^0 \rightarrow J/\psi\, K^{*}(892)^0$. Data is taken from the LHCb experiment collaboration~\cite{LHCb:2013vga}.}
\label{tab:table4}
\end{table}

\subsubsection{$B_d^0 \rightarrow \phi\, K^{*}(892)^0$ }
The polarization amplitudes and the relative phases for the process $B^0 \rightarrow J/\psi K^{*}(892)^0$ are experimentally measured by Belle experiment collaboration~\cite{Belle:2005lvd}. In Table~\ref{tab:table5}, we present the data of polarization amplitudes and relative phases. The correlations of the uncertainties of the polarization amplitudes are not provided by the experimental collaboration~\cite{Belle:2005lvd}. 
\begin{table}[htbp]
\centering
\begin{tabular}{lc}
\toprule
Parameter & Values \\
\midrule
$|A_0|^2$        & $0.45 \pm 0.05  \,(\mathrm{stat.})\pm 0.02(\mathrm{syst.})$ \\
$|A_{\perp}|^2$ & $0.30 \pm 0.06  \,(\mathrm{stat.})\pm 0.02(\mathrm{syst.})$\\
$\delta_{\parallel} \, [\mathrm{rad}]$ & $2.39 \pm 0.24 \,(\mathrm{stat.}) \pm 0.04(\mathrm{syst.}) $\\
$\delta_{\perp} \, [\mathrm{rad}]$    & $2.51 \pm 0.23  \,(\mathrm{stat.})\pm 0.04(\mathrm{stat.})$ \\
\bottomrule
\end{tabular}
\caption{The values of polarization amplitudes and relative phases for $B_d^0 \rightarrow \phi\, K^{*}(892)^0$. Data is taken from the Belle experiment collaboration~\cite{Belle:2005lvd}.}
\label{tab:table5}
\end{table}

\subsubsection{ $B_s^0 \rightarrow J/\psi\, \phi$}
The polarization amplitudes and the relative phases for the process  $B_s^0 \rightarrow J/\psi\, \phi$ are experimentally measured by ATLAS experiment collaboration~\cite{ATLAS:2020lbz}. In Table~\ref{tab:table6}, we present the data of polarization amplitudes and relative phases. Data for correlations in polarization amplitude and relative phases are given in Table~\ref{tab:table7}.
\begin{table}[htbp]
\centering

\begin{tabular}{cc}
\toprule
Parameter & Values \\
\midrule
$|A_0|^2$ & $0.5131 \pm 0.0013\,(\mathrm{stat.}) \pm 0.0038\,(\mathrm{syst.})$ \\
$|A_{\parallel}|^2$ & $0.2213 \pm 0.0019\,(\mathrm{stat.}) \pm 0.0023\,(\mathrm{syst.})$ \\
$\delta_{\parallel} \, [\mathrm{rad}]$ & $3.35 \pm 0.05\,(\mathrm{stat.}) \pm 0.09\,(\mathrm{syst.})$ \\
$\delta_{\perp} \, [\mathrm{rad}]$ & $3.12 \pm 0.11\,(\mathrm{stat.}) \pm 0.06\,(\mathrm{syst.})$ \\
\bottomrule
\end{tabular}
\caption{The values of polarization amplitudes and relative phases for $B_s^0 \rightarrow J/\psi\, \phi$. Data is taken from ATLAS Experiment collaboration~\cite{ATLAS:2020lbz}.}
\label{tab:table6}
\end{table}
\begin{table}[htbp]
\centering
\begin{tabular}{cccccc}
\hline
 & $|A_{\parallel}|^2$ & $|A_0|^2$ & $\delta_{\parallel}$ & $\delta_{\perp}$ \\
\hline
$|A_{\parallel}|^2$ & 1 & -0.341  & 0.522 & 0.133 \\
$|A_0|^2$           &  & 1  & -0.103 & -0.034 \\
$\delta_{\parallel}\,[\mathrm{rad}]$&  &    & 1 & 0.254 \\
$\delta_{\perp}\,[\mathrm{rad}]$    &  &   &  & 1 \\
\hline
\end{tabular}
\caption{Correlation matrix for the polarization amplitudes and phases for $B_s^0 \rightarrow J/\psi\, \phi$. Data is taken from the ATLAS experiment collaboration~\cite{ATLAS:2020lbz}.}
\label{tab:table7}
\end{table}

\begin{figure}[t]
\centering
\includegraphics[width=0.45\textwidth]{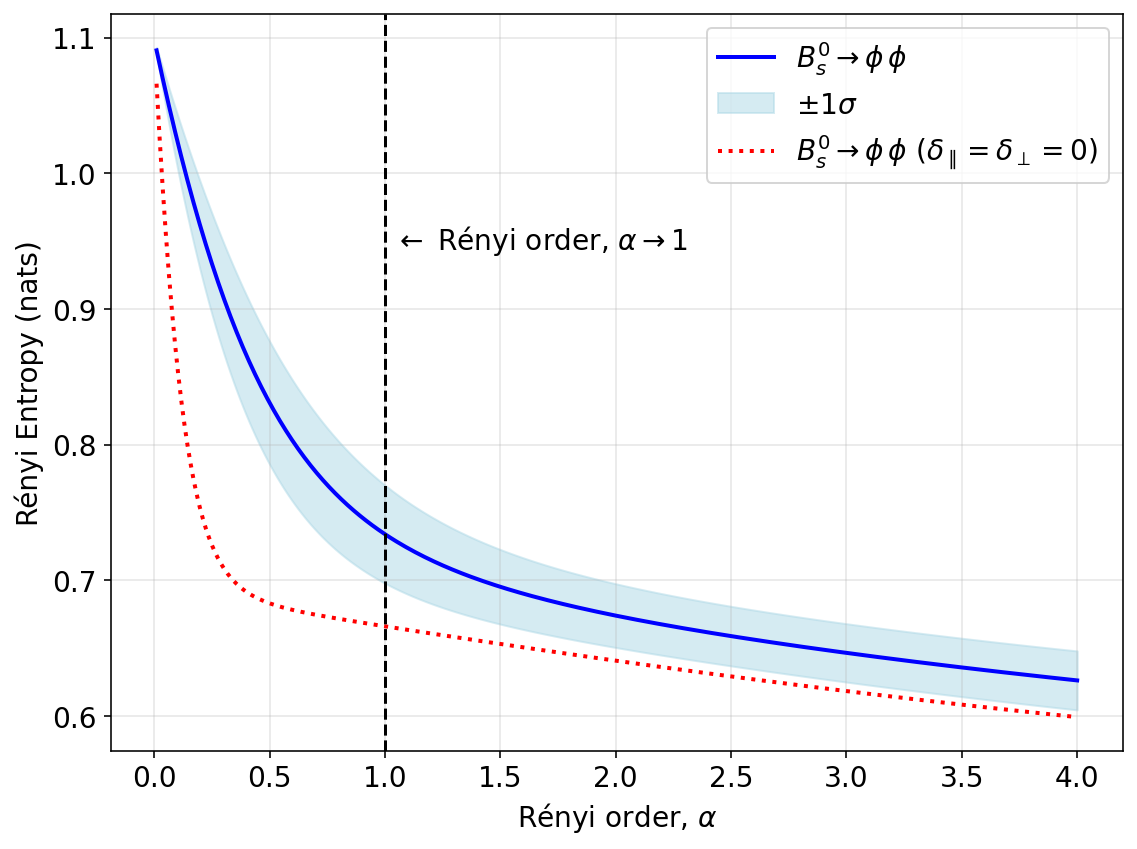}

\caption{(Color online) Plot of the Rényi entropy as a function of Rényi order ($\alpha$) for the process $B_s^0 \rightarrow \phi\, \phi$. Solid line shows the entanglement entropy for non-vanishing phases. Dotted line shows the entanglement entropy for vanishing phases ($\delta_{\parallel} = \delta_{\perp} = 0$). The error band shows the uncertainity in entanglement entropy arises from uncertainity in the measurements of plarization amplitudes and phases. Rényi order, $\alpha \to 1$, line shows the points of von Neumman entropy. }
\label{fig:2}
\end{figure}
\begin{figure}[t]
\centering
\includegraphics[width=0.45\textwidth]{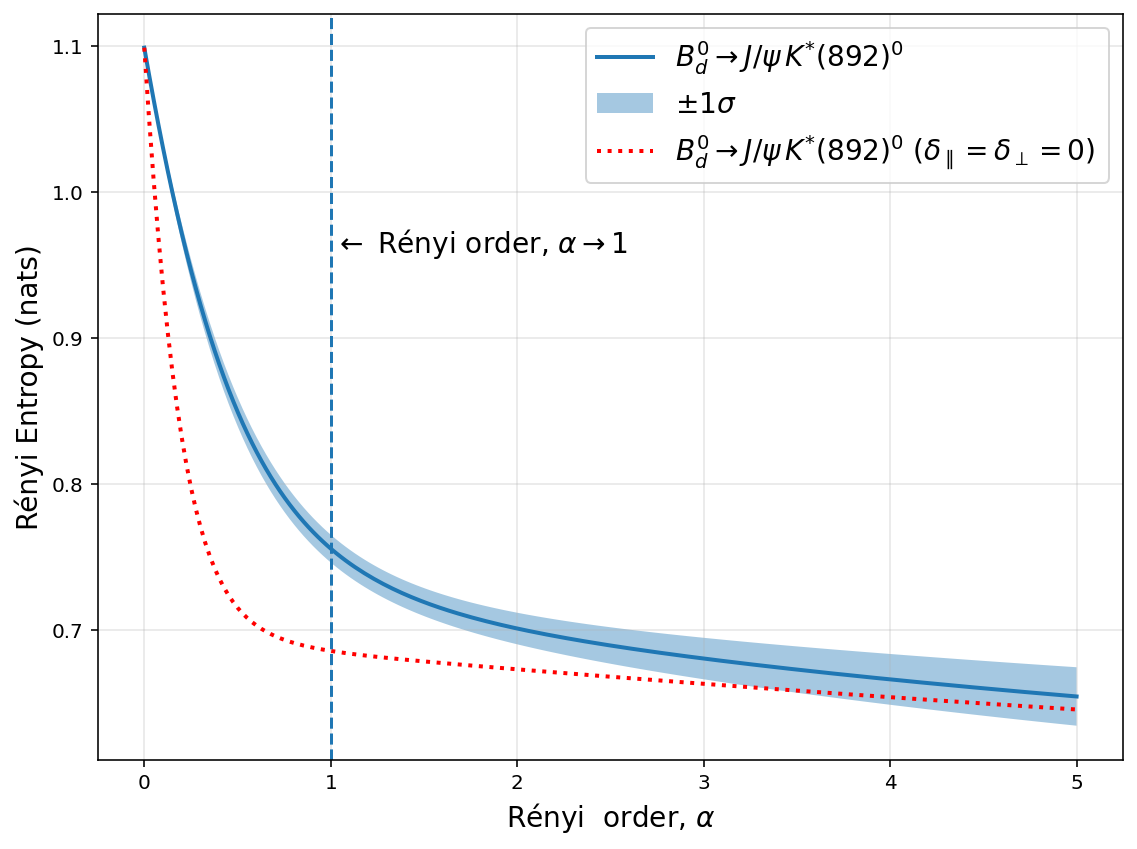}
\caption{(Color online) Plot of Rényi entropy as a function of Rényi order $\alpha$ for the process $B_d^0 \rightarrow J/\psi\, K^{*}(892)^0$ . Solid line shows the entanglement entropy for non-vanishing phases. Dotted line shows the entanglement entropy for vanishing phases ($\delta_{\parallel} = \delta_{\perp} = 0$). The error band shows the uncertainity in entanglement entropy arises from uncertainity in the measurements of plarization amplitudes and phases. Rényi order, $\alpha \to 1$, line shows the points of von Neumman entropy.}
\label{fig:3}
\end{figure}

\begin{figure}[t]
\centering
\includegraphics[width=0.45\textwidth]{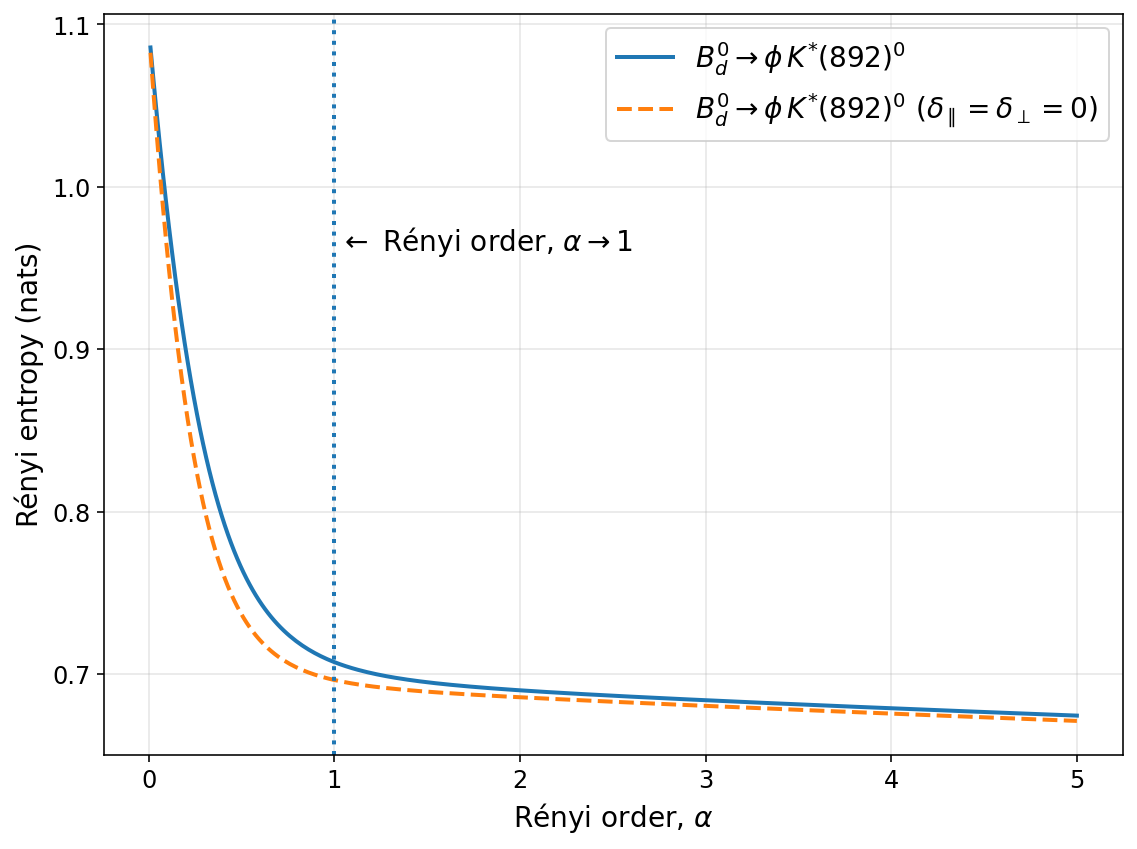}
\caption{(Color online) Rényi entropy as a function of Rényi order ($\alpha$) for the process  $B_d^0 \rightarrow \phi\, K^{*}(892)^0$. Solid line shows the entanglement entropy for non-vanishing phases. Dotted line shows the entanglement entropy for vanishing phases ($\delta_{\parallel} = \delta_{\perp} = 0$). The entropy decreases with increasing ($\alpha$). Rényi order, $\alpha \to 1$, line shows the points of Von Neumman entropy. The correlations of the uncertainties of the polarization amplitudes are not provided by the experimental collaboration~\cite{Belle:2005lvd}. }
\label{fig:4}
\end{figure}

\begin{figure}[t]
\centering
\includegraphics[width=0.45\textwidth]{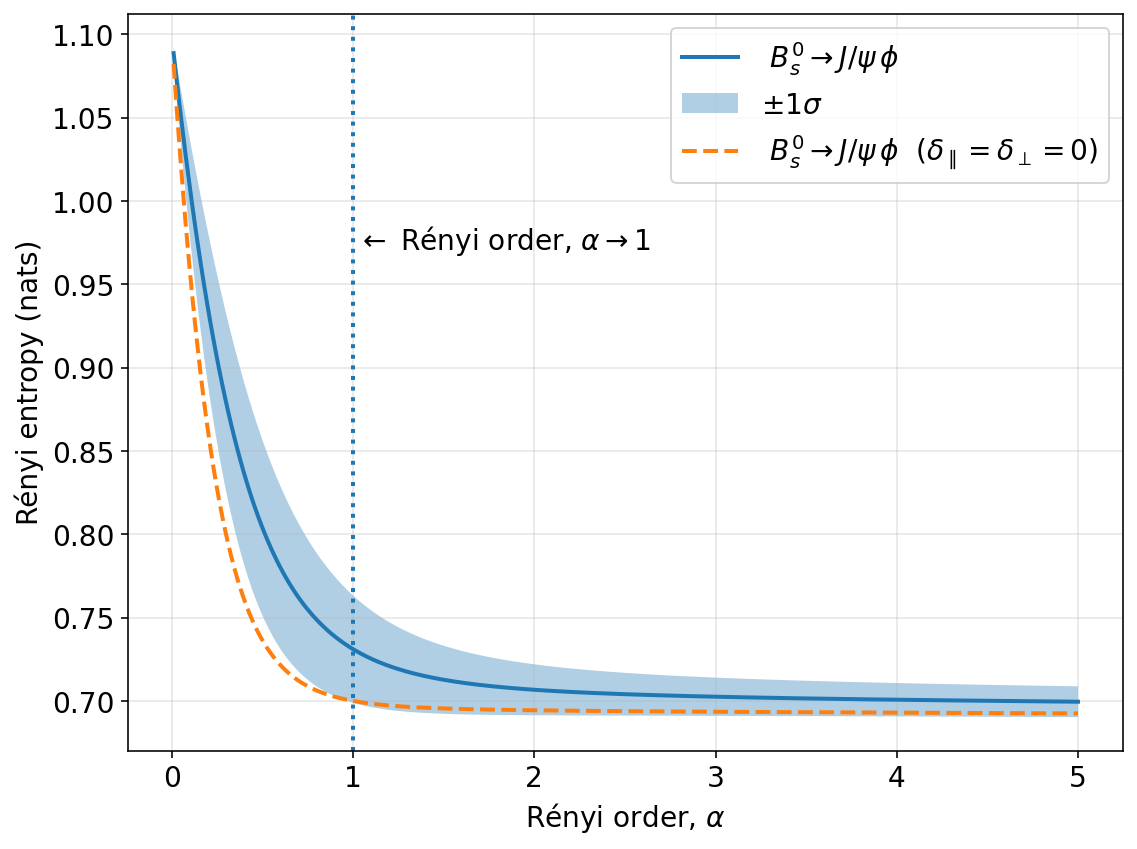}
\caption{(Color online) Plot of Rényi entropy as a function of Rényi order ($\alpha$) for the process  $B_s^0 \rightarrow J/\psi\, \phi$. Solid line shows the entanglement entropy for non-vanishing phases.  Dotted line shows the entanglement entropy for vanishing phases ($\delta_{\parallel} = \delta_{\perp} = 0$). The error band shows the uncertainity in entanglement entropy arises from uncertainity in the measurements of plarization amplitudes and phases. Rényi order, $\alpha \to 1$, line shows the points of Von-Neumman entropy.}
\label{fig:5}
\end{figure}
We have plotted R\'enyi entanglement entropy versus R\'enyi order ($\alpha$) to describe our results. When $\alpha$ approaches 1, we obtain the value of von Neumann entropy. We present the results of entanglement entropy for both vanishing and non-vanishing phase for each process of $B^0$ meson decay in Fig.~\ref{fig:2},~\ref{fig:3},~\ref{fig:4},~\ref{fig:5}. The entanglement entropy is smaller for vanishing phases in comparison to entanglement entropy for non-vanishing phases.

\begin{figure}[t]
\centering
\includegraphics[width=0.45\textwidth]{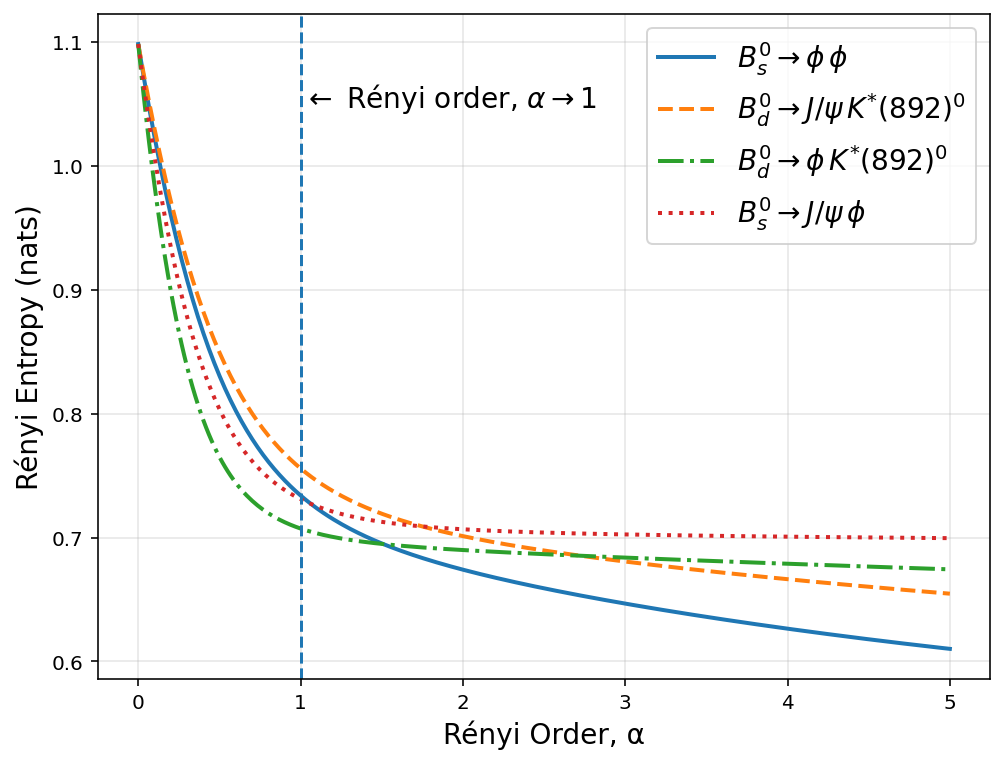}
\caption{(Color online) Comparison of the variation of the Rényi entropy for non-vanishing phases for all the decay process of $B^0$ meson. Solid line shows the variation for $B_s^0 \rightarrow \phi\phi$ process. Dashed line shows the variation for $B_d^0 \rightarrow J/\psi\, K^{*}(892)^0$ process. Dashed-dotted line shows the variation for $B_d^0 \rightarrow \phi\, K^{*}(892)^0$ process. Dotted line shows the variation for  $B_s^0 \rightarrow J/\psi\, \phi$ process. Vertical dashed line for the R\'enyi order $\alpha \to 1$, shows the region of the Von-Neumman entropy.}
\label{fig:6}
\end{figure}
Figure~\ref{fig:6}, shows the comparison of entanglement entropy for different $B^0$ meson decay processes for non-vanishing phases. At small values of $\alpha$ for the process $B_d^0 \rightarrow \phi\, K^{*}(892)^0$ and $B_s^0 \rightarrow J/\psi\, \phi$, the rate of change of the R\'enyi entropy is large. For large values of $\alpha$ for these processes, the rate of change becomes very small and the curve approaches a approximately constant value. 
For small value of $\alpha$ for the process $B_s^0 \rightarrow \phi\ \phi$ and $B_d^0 \rightarrow J/\psi\, K^{*}(892)^0$, the rate of change of the R\'enyi entropy is smaller compared to $B_d^0 \rightarrow \phi\, K^{*}(892)^0$ and $B_s^0 \rightarrow J/\psi\, \phi$. For large values of $\alpha$ for these processes, the rate of change is asymptotically decreases.

We also calculate the error in the entanglement entropy using uncertainty in experimental measurements. The uncertainties in polarization amplitude and phases propagates to the entanglement entropy. We show the error in calculation of entanglement entropy using error band in our plots of R\'enyi entropy versus R\'enyi order ($\alpha$). Error in entanglement entropy is maximum around the value of the Von-Neumman entropy for $B_s^0 \rightarrow \phi\ \phi$ and $B_s^0 \rightarrow J/\psi\, \phi$  processes. Whereas, the error increases for the higher values of $\alpha$ for  $B_d^0 \rightarrow J/\psi\, K^{*}(892)^0$ processes.

\begin{figure}[t]
\centering
\includegraphics[width=0.45\textwidth]{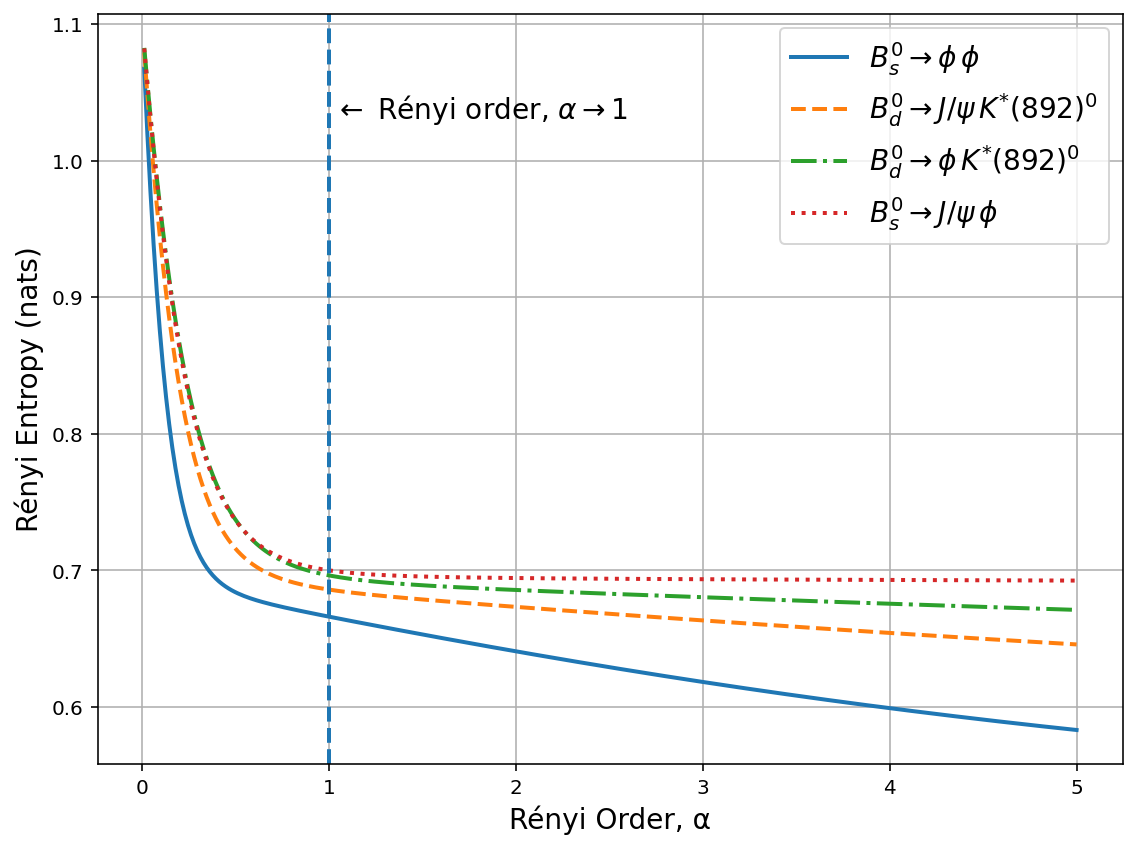}
\caption{(Color online) Comparison of the variation of the Rényi entropy for vanishing phases ($\delta_{\parallel} = \delta_{\perp} = 0$) for all the decay processes of $B^0$ meson. Solid line shows the variation for $B_s^0 \rightarrow \phi\phi$ process. Dashed line shows the variation for $B_d^0 \rightarrow J/\psi\, K^{*}(892)^0$ process. Dashed-dotted line shows the variation for $B_d^0 \rightarrow \phi\, K^{*}(892)^0$ process. Dotted line shows the variation for  $B_s^0 \rightarrow J/\psi\, \phi$ process. Vertical dashed line for R\'enyi order $\alpha \to 1$, shows the region of the Von-Neumman entropy.}
\label{fig:7}
\end{figure}

In the case where phase information is not included, the Rényi entropy depends only on amplitudes. In Figure~\ref{fig:7}, we show the comparison of entanglement entropy for different $B^0$ meson decay processes for vanishing phases ($\delta_{\parallel} = \delta_{\perp} = 0$).  The plot show that there are no intersections between the Renyi entropy curves of different decay processes. Whereas, in the case of non-vanishing phase, we see strong phase dependance and the intersection of the Renyi entropy curves. When phases are zero, the values of entanglement entropy becomes smaller for the same process in comparison with the values for non-zero phases. This shows the importance of phases as they increase the quantum correlations between the particles in the system of massive vector mesons.

\begin{figure}[t]
\centering
\includegraphics[width=0.5\linewidth]{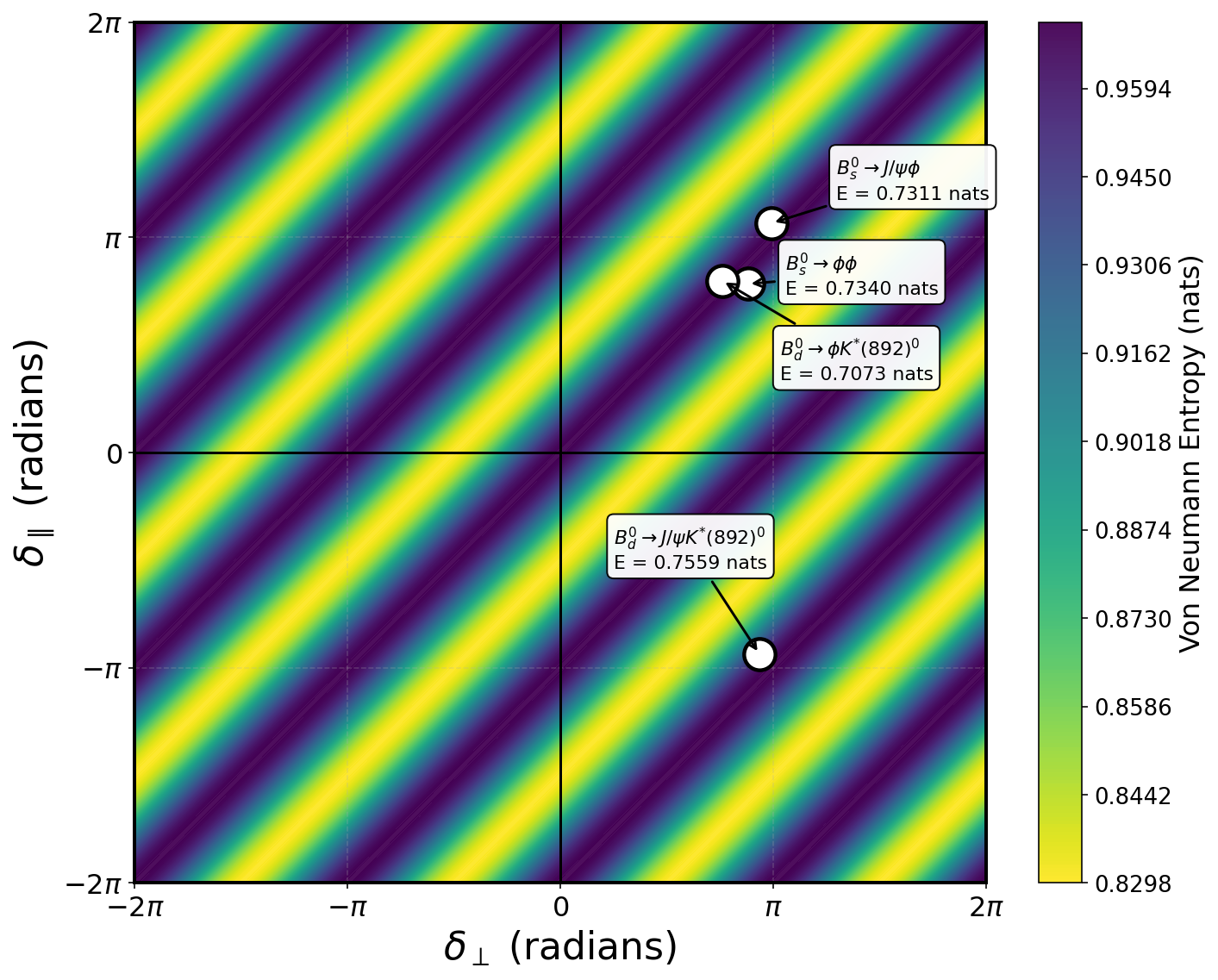}
\caption{(Color online) 2D plot of the phase dependence of the von Neumann entropy for all the processes of $B^0$ meson decay system. The horizontal axis shows $\delta_{\perp}$ and the vertical axis shows $\delta_{\parallel}$, both in radians within the range $[-2\pi,\,2\pi]$. The color bar represents the value of the von Neumman entropy in nats. Yellow regions indicate smaller entropy, while dark purple regions represent larger entropy. The white points represent the von Neumman entropy for the decay channels $B_s^0 \to J/\psi\phi$, $B_s^0 \to \phi\phi$, $B_d^0 \to \phi K^{*}(892)^0$, and $B_d^0 \to J/\psi K^{*}(892)^0$.}
\label{fig:8}
\end{figure}

Figure~\ref{fig:8} shows the 2-D plot for von Neumman entropy dependence on phases. We plot the von Neumann entropy as a function of $\delta_{\perp}$ and $\delta_{\parallel}$. The dark regions show high entropy, which indicates maximum entanglement. The bright regions show lower entropy which indicates minimum entanglement. The Figure~\ref{fig:8} shows a clear diagonal band pattern. These repeating bands indicate that the entropy mainly depends on the phase difference between $\delta_{\perp}$ and $\delta_{\parallel}$ rather than on their individual values. Along the same diagonal band, the entropy remains nearly constant because the relative phase difference does not change. 
We observe that the entanglement entropy is maximum for the $B_d^0 \to J/\psi\, K^{*}(892)^0$ process, while it is minimum for the $B_s^0 \to J/\psi\, \phi$ process. The entanglement entropy values for $B_d^0 \to J/\psi\, K^{*}(892)^0$ is on the off-diagonal band whereas the values of entropy for other process lies on the diagonal band in the 2D plots.
 
\begin{table*}[htbp]
\centering
\small
\begin{tabular}{ c c c c c }
\hline
\makecell{$B^0$ meson decay\\ processes}
& \makecell{$\alpha \to 0$ \\ (Hartley entropy,\\ Max-entropy)}
& \makecell{$\alpha \to 1$ \\ (Von Neumann\\ entropy)}
& \makecell{$\alpha = 2$ \\ (Collision entropy)}
& \makecell{$\alpha \to \infty$ \\ (Min-entropy)} \\
\hline

$B_s^0 \rightarrow \phi\,\phi$
& $1.098612$
& $0.733984 \pm 0.036125$
& $0.673959 \pm 0.023487$
& $0.508328 \pm 0.020969$ \\
\hline

$B_d^0 \rightarrow J/\psi\,K^{*}(892)^0$
& $1.098612$
& $0.755890 \pm 0.009491$
& $0.701210 \pm 0.010806$
& $0.558616 \pm 0.025759$ \\
\hline

$B_d^0 \rightarrow \phi\,K^{*}(892)^0$
& $1.098612$
& $0.707335$
& $0.689956$
& $0.603508$ \\
\hline

$B_s^0 \rightarrow J/\psi\,\phi$
& $1.098612$
& $0.731103 \pm 0.032203$
& $0.706735 \pm 0.015286$
& $0.667285 \pm 0.007796$ \\
\hline
\end{tabular}
\caption{Rényi entropy values for different Rényi orders ($\alpha$) for non-vanishing strong phases. The von Neumann entropy corresponds to the limit $\alpha \to 1$.}
\label{tab:Table8}
\end{table*}


\begin{table*}[htbp]
\centering
\begin{tabular}{ c c c c c c c }
\hline
\makecell{$B^0$ meson decay\\ processes}
&\makecell{$\alpha\to0$ \\ ( Hartley entropy,\\ Maximum entropy)}
&\makecell{$\alpha \to 1$ \\ (Von Neumann\\ entropy)}  
&\makecell{$\alpha = 2$\\ (Collision entropy)} 
&\makecell{$\alpha \to \infty$ \\(Minimum entropy)}\\
\hline

$B_s^0 \rightarrow \phi\, \phi$
& 1.098612
&  0.666109
&  0.640745
&  0.484519\\

$B_d^0 \rightarrow J/\psi\, K^{*}(892)^0$ 
&  1.098612
&  0.685933
&  0.673294
&  0.558616\\

$B_d^0 \rightarrow \phi\, K^{*}(892)^0$ 
&  1.098612
&  0.696314
&  0.685675
&  0.599910\\ 

$B_s^0 \rightarrow J/\psi\, \phi$  
&  1.098612
&  0.700053 
&  0.694425
&  0.667285 \\

\hline
\end{tabular}
\caption{Rényi entropy values for different Rényi order ($\alpha$)  for vanishing phases ($\delta_{\parallel} = \delta_{\perp} = 0$). The von Neumman entropy corresponds to limit of $\alpha \to 1$.}
\label{tab:Table9}
\end{table*}

In the end, we list the results for the entropy of entanglement for different Rényi orders ($\alpha$) in the tabular form. In Table~\ref{tab:Table8} and Table~\ref{tab:Table9}, we present the value of entanglement entropy for vanishing and non vanishing phases. At $\alpha$ = 0, the value of entanglement entropy is maximum and constant for all processes. We show that entanglement entropy is large for non-vanishing phases in comparison to the entanglement entropy for vanishing phases. 

In Table~\ref{tab:Table10}, we present the result of von Neumman entropy along with errors in entropy for both vanishing and non-vanishing phases. We compare the results of the von Neumman entropy for vanishing and non-vanishing phases in Table~\ref{tab:Table10}. The von Neumann entropy is maximum for $B_d^0 \rightarrow J/\psi\, K^{*}(892)^0$ for non vanishing phases while it is maximum for $B_s^0 \rightarrow J/\psi\, \phi$ for vanishing phases. Also, it is a monotonically increases from $B_s^0 \rightarrow \phi\, \phi$ to $B_s^0 \rightarrow J/\psi\, \phi$ for the vanishing phase.
The results of the von Neumann entropy show a larger error for the $B_s^0 \rightarrow J/\psi\, \phi$ and $B_s^0 \rightarrow \phi\, \phi$ processes compared to the error in von Neumann entropy for $B_d^0 \rightarrow J/\psi\, K^{*}(892)^0$ process.

\begin{table*}[htbp]
\centering
\begin{tabular}{ c c c c c }
\hline
$B^0$ meson decay processes & Relative phases & \makecell{Von Neumann entropy \\ ($\alpha \to 1$)} \\
\hline

$B_s^0 \rightarrow \phi\, \phi$ & 

\makecell{$\delta_{\parallel}=2.463,\ \delta_{\perp}=2.769$ \\ 
$\delta_{\parallel} = \delta_{\perp} = 0$} & 
\makecell{$ 0.733984 \pm 0.036125$ \\ $ 0.666069$}  \\
\hline

$B_d^0 \rightarrow J/\psi\, K^{*}(892)^0$ & 
\makecell{$\delta_{\parallel}=-2.94,\ \delta_{\perp}=2.94$ \\ 
$\delta_{\parallel} = \delta_{\perp} = 0$} & 
\makecell{$ 0.755899 \pm 0.009491$ \\ $0.685933$}  \\
\hline

$B_d^0 \rightarrow \phi\, K^{*}(892)^0$ & 
\makecell{$\delta_{\parallel}=2.72,\ \delta_{\perp}=2.81$ \\ 
$\delta_{\parallel} = \delta_{\perp} = 0$} & 
\makecell{$0.707335$ \\ $0.696314$} \\
\hline

$B_s^0 \rightarrow J/\psi\, \phi$  & 
\makecell{$\delta_{\parallel}=3.35,\ \delta_{\perp}=3.12$ \\ 
$\delta_{\parallel} = \delta_{\perp} = 0$} & 
\makecell{$0.731103 \pm 0.032203$ \\ $0.700053$} \\
\hline

\end{tabular}

\caption{The von Neumann entropy for the decay of $B^0$ mesons for vanishing and non-vanishing phases. Data for relative phases are taken from \cite{Gabrielli:2024kbz,LHCb:2013vga,Belle:2005lvd,ATLAS:2020lbz}.}
\label{tab:Table10}
\end{table*}

\begin{table*}[htbp]
\centering
\begin{tabular}{c c c }
\hline
\makecell{$B^0$ meson decay processes} & 
\makecell{Branching \\ fraction} & 
\makecell{Von Neumann entropy \\ ($\alpha \to 1$)} \\
\hline

$B_d^0 \rightarrow J/\psi\, K^{*}(892)^0$ & 
$(1.27 \pm 0.05)\times 10^{-3}$ &
$0.755899 \pm 0.009491$ \\ 
\hline

$B_s^0 \rightarrow J/\psi\, \phi$  & 
$(1.04 \pm 0.04)\times 10^{-3}$ &
$0.731103 \pm 0.032203$ \\
\hline

$B_s^0 \rightarrow \phi\, \phi$ & 
$(1.85 \pm 0.14)\times 10^{-5}$ &
$0.733984 \pm 0.036125$ \\ 
\hline

$B_d^0 \rightarrow \phi\, K^{*}(892)^0$ & 
$(1.00 \pm 0.05)\times 10^{-5}$ &
$0.707335$ \\ 
\hline
\end{tabular}
\caption{The von Neumman entropy and the Branching fractions for the decay of $B^0$ meson. The data of branching fraction are taken from particle data group~\cite{ParticleDataGroup:2026aaa}.}
\label{tab:table11}
\end{table*}

\begin{table*}[htbp]
\centering
\caption{Linear entropy $(S_L)$, I-concurrence $(C_I)$, Tangle $(\tau)$, Negativity $(\mathcal{N})$, Logarithmic Negativity $(E_{\mathcal{N}})$, Schmidt coefficients, and Schmidt rank for different $B^0$ meson decay processes.}
\label{tab:table12}
\small 
\setlength{\tabcolsep}{0pt} 
\begin{tabular*}{\textwidth}{@{\extracolsep{\fill}} l c  c c c c c }
\toprule
\textbf{Decay process} 
& \textbf{$S_L$} 
& \textbf{$C_I$} 
& \textbf{$\tau$}
& \textbf{$\mathcal{N}$}
& \textbf{$E_{\mathcal{N}}$}
& \begin{tabular}[c]{@{}c@{}}\textbf{Schmidt coefficients}\\ \textbf{(Schmidt rank = 3)}\end{tabular} \\
\midrule

$B_s^0 \rightarrow \phi\phi$
& 0.4906
& 0.9906
& 0.9812
& 0.64765
& 0.83087
& $(0.1196,\;0.6198,\;0.7756)$ \\

$B_d^0 \rightarrow J/\psi\, K^{*}(892)^0$
& 0.5040
& 1.0040
& 1.0080
& 0.66953
& 0.84981
& $(0.7563,\;0.6407,\;0.1324)$  \\

$B_d^0 \rightarrow \phi\, K^{*}(892)^0$ 
& 0.4975
& 0.9975
& 0.9950
& 0.57520
& 0.76565
& $(0.0558,\;0.6708,\;0.7395)$  \\

$B_s^0 \rightarrow J/\psi \phi$
& 0.5068
& 1.0067
& 1.0135
& 0.61710
& 0.80389
& $(0.0860,\;0.6925,\;0.7163)$ \\

\bottomrule
\end{tabular*}
\end{table*}

In Table~\ref{tab:table11}, we compare the branching fractions of different decay processes with the value of von Neumann entropy. The comparison reveals that the von Neumann entropy is larger for $B^0$ meson decay channels with larger branching fractions. This suggests a direct connection between quantum entanglement and the decay dynamics involving strong and weak interactions. A larger branching fraction indicate that the decay process occurs more frequently, which shows a stronger transition from the initial $B^0$ meson state to the final two vector mesons state. 
From the results, we can notice that the \(B_d^0 \rightarrow J/\psi\, K^{*}(892)^0\) decay process has the largest branching fraction among all processes. This decay process also gives the largest value of the von Neumann entropy. This indicates that the produced final state has a comparatively stronger quantum correlations. 
The decay process \(B_d^0 \rightarrow \phi\, K^{*}(892)^0\) gives the smallest von Neumann entropy among all processes. This indicates that the process \(B_d^0 \rightarrow \phi\, K^{*}(892)^0\) shows relatively weaker quantum correlations in the final state. A relation between branching fractions and entanglement entropy, if established using phenomenological studies, will be useful for using entanglement entropy as a measure for understanding decay dynamics.

Table~\ref{tab:table12} shows the different entanglement measures for several $B^0$ meson decay processes. The table includes linear entropy $(S_L)$, I-concurrence $(C_I)$, tangle $(\tau)$, negativity $(\mathcal{N})$, logarithmic negativity $(E_{\mathcal{N}})$, Schmidt coefficients, and the Schmidt rank. The Linear entropy $(S_L)$, I-concurrence $(C_I)$ and tangle $(\tau)$ are maximum for the process $B_s^0 \rightarrow J/\psi \phi$ and minimum for the processes \(B_s^0 \rightarrow \phi\, \phi\), whereas negativity $(\mathcal{N})$ and logarithmic negativity $(E_{\mathcal{N}})$ are maximum for the process $B_d^0 \rightarrow J/\psi\, K^{*}(892)^0$ and minimum for the processes \(B_d^0 \rightarrow \phi\, K^{*}(892)^0\). Together, these quantities help us to understand the strength of the quantum entanglement in the final vector meson systems.

The values of linear entropy for all processes are close to $0.5$, while the maximum value of linear entropy for two qutrit system is 0.6666. This indicates that the reduced density matrices are strongly mixed and the two vector mesons are highly quantum correlated. Among all processes, $B_s^0 \rightarrow J/\psi \phi$ gives the largest linear entropy, which suggests slightly stronger mixedness in this decay channel. 

The I-concurrence and tangle values are close to unity for all processes, while the maximum value of I-concurrence for two qutrit system is 1.1547. Since I-concurrence measures the strength of entanglement, values near one indicate highly entangled states. The tangle values also confirm the presence of strong quantum correlations. The process $B_s^0 \rightarrow J/\psi \phi$ gives the largest I-concurrence and tangle values among the considered decay modes.

The negativity and logarithmic negativity are also nonzero for all processes, which clearly confirms the existence of quantum entanglement. The largest values of negativity and logarithmic negativity are 0.66953 and 0.84981 respectively for the process $B_d^0 \rightarrow J/\psi\, K^{*}(892)^0$, while the maximum limiting value for negativity and logarithmic negativity are 1 and 1.098612 respectively. Among all channels, $B_d^0 \rightarrow J/\psi\, K^{*}(892)^0$ gives the largest negativity and logarithmic negativity values, indicating comparatively stronger correlations.

The Schmidt coefficients describe how the quantum state is distributed among different helicity basis states. In every process, more than one Schmidt coefficient has a significant contribution. This shows that the entanglement is shared among multiple helicity states rather than coming from a single dominant configuration. The Schmidt rank is $3$ for all decay channels, which means that all three helicity basis states participate in the entangled system.

Overall, the table shows that all studied $B^0$ meson decay processes exhibit strong quantum entanglement. All channels display significant nonclassical quantum correlations. These results support the idea that heavy meson decay systems can provide an important platform for studying quantum information properties in high-energy particle physics.

\section{Summary and Outlook}
In this work, we have presented the results of the quantum correlation study for the decay of the $B^0$ meson into a system of two massive vector mesons. We present a phenomenological study for $B_s^0 \rightarrow \phi\, \phi$, $B_d^0 \rightarrow J/\psi\, K^{*}(892)^0$, $B_d^0 \rightarrow \phi\, K^{*}(892)^0$ and $B_s^0 \rightarrow J/\psi\, \phi$ where we calculate various measures of entanglement. The Decay of spin-0 $B^0$ meson to two massive spin-1 particles forms a bipartite system of entangled qutrits. As the bipartite system is in the pure state, we calculate the various entanglement measures using the density matrix formalism. The experimental measurements of polarization amplitudes and relative phases of decaying particles allowed us to write the entangled state of the two massive vector mesons. We present the result of entanglement entropy and various other measures of quantum entanglement. The R\'enyi entropy is the gneralized form of the von Neumann entropy, an important measure of quantum entanglement. The different values and limit of the R\'enyi order ($\alpha$) provide different interpretations of entropy and entanglement in the system. The results of entanglement entropy at different values and limits of $\alpha$ provide an important insight related to Hartley entropy (Max-entropy), min-entropy, collision entropy, and the von Neumann entropy. These interpretations are crucial for understanding the quantum state of the two massive vector mesons. We also study the effect of the relative phase on the variation of R\'enyi entropy. Our results show a strong phase dependence on the R\'enyi entropy at different values and limits of $\alpha$. The entanglement entropy and its dependence on relative phase may provide critical insight into the decay dynamics and CP violation related studies in the $B^0$ meson decay. The results show that von Neumann entropy is larger for decay processes of $B^0$ meson with larger branching fraction, which indicate the crucial relation of quantum entanglement with decay dynamics involving strong and weak interaction. 
A relation between branching fractions and entanglement entropy, if established using phenomenological studies, will be useful for using entanglement entropy as a measure for understanding decay dynamics.
Our analysis show that the value of von Neumann entropy is maximum for $B_d^0 \rightarrow J/\psi\, K^{*}(892)^0$ with minimum error when compared with values of other decay processes. In addition to study of entanglement entropy, we present results of linear entropy $(S_L)$, I-concurrence $(C_I)$, tangle $(\tau)$, negativity $(\mathcal{N})$, logarithmic negativity $(E_{\mathcal{N}})$, Schmidt coefficients, and Schmidt rank for different $B^0$ meson decay processes. Our results show that the linear entropy $(S_L)$, I-concurrence $(C_I)$, tangle $(\tau)$ is the maximum for the decay process $B_s^0 \rightarrow J/\psi\, \phi$. These measures provide degree of quantum entanglement and information about the purity of the system. The maximum value of negativity $(\mathcal{N})$ and logarithmic negativity $(E_{\mathcal{N}})$ is corresponds to the decay process $B_d^0 \rightarrow J/\psi\, K^{*}(892)^0$. The entanglement measures related to decay of the $B^0$ meson may become critical to understanding the decay dynamics and studies related to CP violation in high energy physics. 

\begin{acknowledgments}
The authors thank Prof. Anirban Pathak for their valuable guidance, insightful discussions, and constructive feedback throughout the development of this work.
\end{acknowledgments}

\appendix






\nocite{*}
\bibliography{apssamp}

\end{document}